\documentclass[aps,prl,twocolumn,twoside,nofootinbib,preprintnumbers,floatfix,longbibliography]{revtex4-2}
\usepackage{hyperref}
\usepackage{graphicx}
\usepackage{amsmath}
\usepackage{mathrsfs}
\usepackage{xspace}
\usepackage{amsfonts}
\usepackage{color}   
\usepackage{booktabs,siunitx}
\newcommand{\eq}[1]{Eq.~\eqref{eq:#1}}
\newcommand{\eqs}[2]{Eqs.~\eqref{eq:#1} and \eqref{eq:#2}}
\newcommand{\fig}[1]{Fig.~\ref{fig:#1}}

\newcommand{\nn}{\nonumber}

\allowdisplaybreaks[2]

\usepackage{xcolor}
\usepackage{marginnote}
\usepackage[normalem]{ulem}
\usepackage{dsfont,mathrsfs}

\definecolor{darkgreen}{rgb}{0.13,0.55,0.13}
\definecolor{or}{rgb}{0.88,0.43,0.02}

\begin{document}
\title{Dihadron Fragmentation and the Confinement Transition in Energy Correlators}

\author{Kyle Lee}%
\email{kylel@mit.edu}%
\affiliation{Center for Theoretical Physics -- a Leinweber Institute, Massachusetts Institute of Technology, Cambridge, MA 02139, USA}

\author{Iain W. Stewart}%
\email{iains@mit.edu}%
\affiliation{Center for Theoretical Physics -- a Leinweber Institute, Massachusetts Institute of Technology, Cambridge, MA 02139, USA}

\begin{abstract}
In this letter, we relate the factorization for $e^+e^- \to h_1 h_2 X$ to the factorization for energy-energy correlators in the collinear limit. This enables us to give a nonperturbative proof of factorization for the energy correlators, 
relate the energy correlator jet function to  transverse-momentum-sensitive dihadron fragmentation functions, and provide a rigorous description of the confinement transition region.
\end{abstract}

\preprint{MIT-CTP 5889}

\maketitle

\emph{\textbf{Introduction.}} In high-energy collider experiments, the dynamics of Quantum Chromodynamics (QCD) are encoded in the energy distributions of hadrons recorded by detectors. A central puzzle is how the ultraviolet partons produced in hard scattering evolve into the observed infrared hadrons, and precisely how this confinement leaves its imprint on observables. Energy-weighted correlation functions of hadrons, known as energy correlators~\cite{Basham:1979gh,Basham:1978zq,Basham:1978bw,Basham:1977iq}, have recently been identified as powerful probes of hadronization and confinement dynamics. For a recent review, see~\cite{Moult:2025nhu}.

The simplest non-trivial correlator is between two hadrons, also known as energy-energy correlators (EEC), $\left\langle\Psi\left|\mathcal{E}\left(\vec{n}_1\right) \mathcal{E}\left(\vec{n}_2\right)\right| \Psi\right\rangle$, where $|\Psi\rangle = \mathcal{O}|0\rangle$ represents a state generated by a source operator $\mathcal{O}$ and the energy flow operator gives hadronic energies along the direction of the energy flow operator as $\mathcal{E}(\vec{n})|X\rangle = \sum_h E_h \delta^2(\Omega_{\vec{p}_h}-\Omega_{\vec{n}})|X\rangle$. For example, in $e^+e^- \to \gamma^* \to$ hadrons, the electromagnetic current $J^\mu= e_f\bar{\psi}_f\gamma^\mu\psi_f$ can source a state with definite energy and momentum $q=l'-l$ with the lepton momenta $l^\mu$ and $l'^\mu$ as
\begin{align}
\label{eq:EECdef}
\langle&\mathcal{E}\left(\vec{n}_1\right) \mathcal{E}\left(\vec{n}_2\right)\rangle_q\nonumber \\
& \equiv \frac{1}{\sigma}\int \mathrm{~d}^4 x \frac{e^{i q\cdot x}}{Q^2}L_{\mu\nu}\langle 0| J^{\mu \dagger}(x) \mathcal{E}\left(\vec{n}_1\right) \mathcal{E}\left(\vec{n}_2\right) J^\nu(0)|0\rangle\nonumber\\
&= \frac{1}{\sigma_{\rm tot}}\sum_X\sum_{h_i,h_j}\int d \sigma_{e^{+} e^{-} \rightarrow h_i h_jX}  \frac{E_{h_i}E_{h_j}}{Q^2}\nonumber\\
&\quad \times \delta^2(\Omega_{\vec{p}_{h_i}}-\Omega_{\vec{n}_1}) \delta^2(\Omega_{\vec{p}_{h_j}}-\Omega_{\vec{n}_2})\,,
\end{align}
where $Q=\sqrt{q^2}$ is the center-of-mass energy of the $e^+e^-$, $L^{\mu\nu} = l^\mu l'^\nu + l'^\mu l^\nu - (l\cdot l') g^{\mu\nu}$ is the leptonic tensor, $\sigma =L_{\mu\nu}\times\int d^4 x\, e^{i q\cdot x}\langle 0| J^{\mu\dagger}(x) J^\nu(0)|0\rangle$,  and $\sigma_{\rm tot} =\sigma\times (4\pi \alpha_{\rm EM})^2/2Q^6$ is the total cross section. 

Measuring only the angle between the two detectors via $z = (1-\vec{n}_1\cdot \vec{n}_2)/2$ gives the EEC,
\begin{align}
\label{eq:EECdef2}
\frac{1}{\sigma_{\rm tot}}\frac{d\sigma_{\rm EEC}}{dz} \equiv& \int d^2\Omega_{\vec{n}_1}d^2\Omega_{\vec{n}_2}\\
&\hspace{-1cm}\times\delta\left(z-\frac{1-\vec{n}_1\cdot \vec{n}_2}{2}\right)\langle\mathcal{E}\left(\vec{n}_1\right) \mathcal{E}\left(\vec{n}_2\right)\rangle_q\,\,.\nonumber
\end{align} 
As is clear from~\eq{EECdef}, the EEC of hadrons is directly related to dihadron production $e^+e^- \to h_i h_j X$ at various angles. This connection was, for example, leveraged to derive the EEC factorization in the back-to-back region ($1-z\ll 1$)~\cite{Moult:2018jzp}. 
On the other hand, the connection between dihadron production and the EEC in the collinear region ($z\ll 1$) 
has not been harnessed. In this letter, we explore this connection to provide a nonperturbative proof of the EEC collinear factorization and derive an explicit formula for the EEC jet function in terms of dihadron fragmentation functions, which enables us to make rigorous QCD-based predictions for the description of the confinement transition.

The EEC is an infrared safe observable that is perturbatively computable as long as its invariant-mass scale $Q\sqrt{z(1-z)}\gg \Lambda_{\rm QCD}$. 
High precision fixed-order calculations~\cite{Dixon:2018qgp,Tulipant:2017ybb}, and the resummation of twist-2 and Sudakov double logarithms in the collinear and back-to-back regions~\cite{Konishi:1979cb,Collins:1981zc,Moult:2018jzp,Dixon:2019uzg,Chen:2023zzh,Duhr:2022yyp}, respectively, make the EEC one of the most precisely understood observables.

\begin{figure}[!t]
\begin{center}
\includegraphics[width=0.44\textwidth]{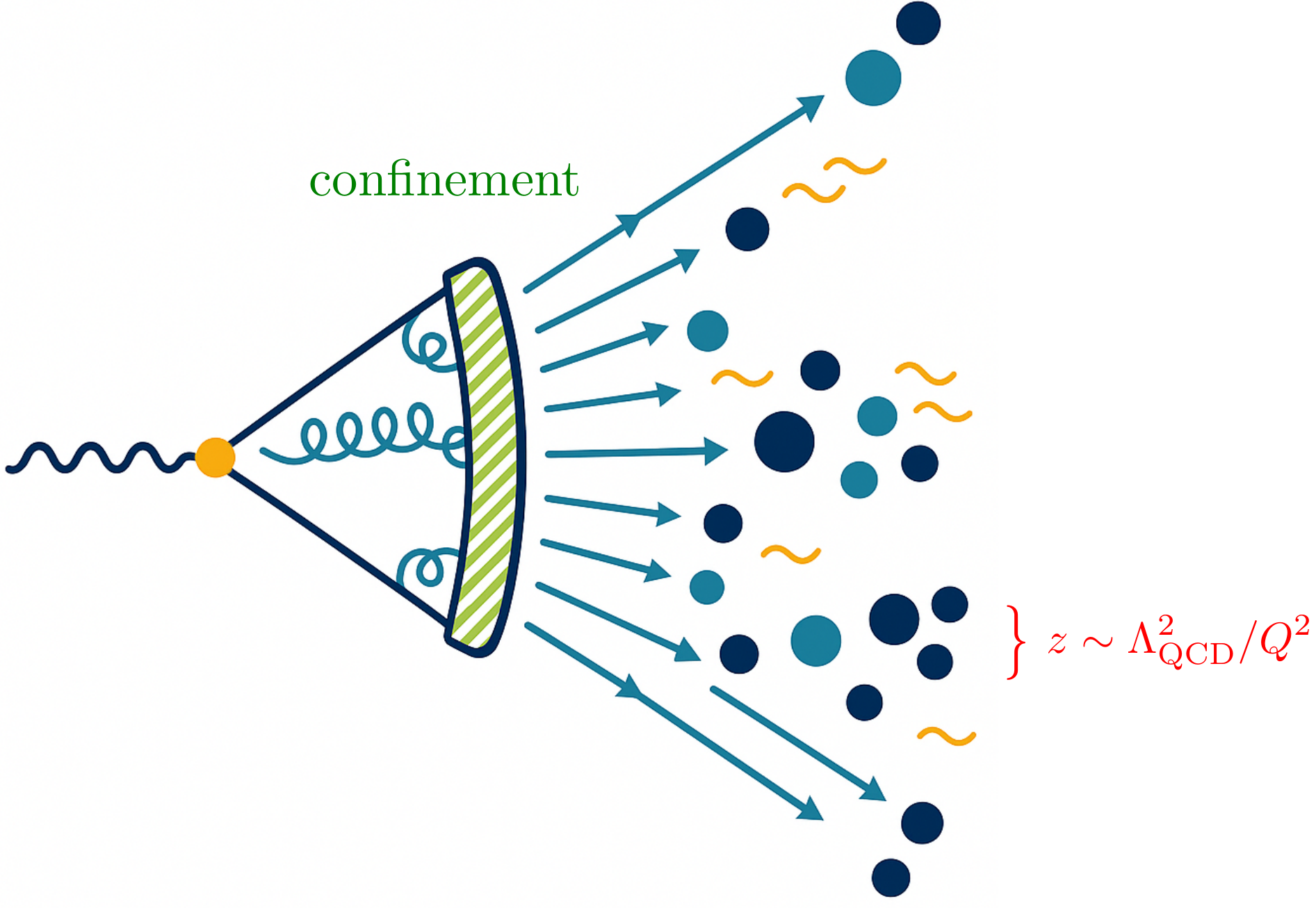}
\end{center}
\caption{Partons from a hard scattering undergo a cascade of partonic splittings at small angles $1\gg z\gg \Lambda_{\rm QCD}^2/Q^2$, and ultimately confine via hadronization into a dense cloud of uncorrelated hadrons separated by $z \sim \Lambda_{\rm QCD}^2/Q^2$.}
\label{fig:confinement}
\end{figure}

Near the $z\to 0,1$ endpoints, the invariant mass scale becomes nonperturbative, $Q\sqrt{z(1-z)} \sim \Lambda_{\rm QCD}$. 
For $z\to1$, perturbative Sudakov logarithms and confinement both strongly suppress  $d\sigma_{\rm EEC}/dz$, making it hard to disentangle their individual effects.
In contrast, in the small-angle regime, there is a clear signature of confinement with a transition from a regime governed by the scaling of twist-2 operators dressing a $\sim 1/z$ singularity to one of vanishing correlations among hadrons.
At very small angles, nonperturbative confinement dynamics generate a dense cloud of hadrons  as illustrated in~\fig{confinement}, and
wash out perturbative correlations. 
This parton-to-hadron transition has recently been observed experimentally~\cite{STAR:2025jut,ALICE:2024dfl,CMS:2024mlf} and sets confining theories such as QCD apart from conformal theories, where the light-ray OPE scaling~\cite{Chen:2021gdk,Hofman:2008ar,Kravchuk:2018htv,Kologlu:2019mfz} persists to arbitrarily small angles. This striking feature of confinement has been studied in many recent publications~\cite{Liu:2024lxy,Barata:2024wsu,Csaki:2024zig,Csaki:2024joe,Lee:2024jnt}, and one of our main goals is to provide a description of this transition from first principles.

One of the central organizing principles in high-energy hadronic processes is ``QCD factorization'', which achieves predictive power by separating dynamics at disparate scales and isolating universal, nonperturbative functions that can be extracted in one kinematic regime and applied in another. Guided by this principle, it is then natural to ask: what is the universal nonperturbative structure underlying the small-angle confinement transition observed in the EEC, and how can we cleanly disentangle perturbative from genuine nonperturbative effects? In this letter, we address these questions by connecting the EEC and dihadron production in the collinear limit. This allows us to derive the all-order factorization of the EEC in the full collinear region $(z\ll 1)$ valid both in the OPE region ($\Lambda_{\rm QCD}^2/Q^2\ll z\ll 1$) and in the confinement transition region ($Q^2z\sim \Lambda_{\rm QCD}^2$), define universal nonperturbative jet functions 
that fully capture the confinement transition, and derive their perturbative evolution to enable predictive power.

\emph{\textbf{Dihadron Production.}} In collinear dihadron production, both hadrons emerge from the same fragmenting parton. Accordingly, the relevant nonperturbative functions describing such fragmenting process are the transverse-momentum-sensitive dihadron fragmentation functions (FFs)~\cite{Pitonyak:2025lin,Pitonyak:2023gjx,Rogers:2024nhb,Jaffe:1997hf,Bianconi:1999cd}. For a quark, they are~\cite{Pitonyak:2025lin,Pitonyak:2023gjx}
\begin{align}
\label{eq:dihadQk}
& D_{q\to h_j h_k}(y_j,y_k,\vec{P}_{j\perp},\vec{P}_{k\perp})=
\! \int\!\! \frac{d r^{+} d^2 \vec{r}_{\perp}}{8N_c y_j y_k (2 \pi)^6} \, e^{i k^- r^+}\sum_X \hspace{-0.5cm} 
\int \nn \\
&\times \operatorname{Tr}\bigl[\gamma^- \bigr\langle 0|  \psi_q(r^+,0,r_\perp)\bigl|h_j,h_k; X\bigr\rangle\bigl\langle h_j,h_k; X\bigr| \bar{\psi}_q(0)|0\rangle\bigr] ,
\end{align}
where $\sum_X \hspace{-0.76cm}$ \large$\int$\normalsize$\,\ \ \equiv  \ \!\! \sum_X \prod_{i\in X} \int\!\! \frac{d^3 \vec{p}_i}{(2\pi)^3 \, 2 p_i^0}$,
and similarly for a gluon. We use the convention $v^{\pm}\equiv (v^0\pm v^3)/\sqrt{2}$ and suppress gauge links for convenience. The variable $y_a=p_{h_a}^-/k^-$ is the longitudinal momentum fraction of hadron $h_a$ with respect to the fragmenting parton and $\vec{P}_{a\perp}$ is the transverse momentum of $h_a$ in the frame where  this parton has no transverse momentum ($a=j,k$).

\begin{widetext}
In the region $\{ \Lambda_{\rm QCD}^2, \vec{P}_{a\perp}^2 \}\ll Q^2$, the two hadrons originate from a single fragmenting parton, and hence have the same $H\otimes D$ factorization~\cite{Pitonyak:2025lin,Pitonyak:2023gjx,Rogers:2024nhb} as the single-inclusive hadron production, with the same hard coefficients $H_i(x,\frac{Q}{\mu},\mu)$,
\begin{align}
\label{eq:dihadprod}
\frac{1}{\sigma_0}\frac{d\sigma^{h_j h_kX}}{ d\xi_j d\xi_k d^2\vec{P}_{j\perp}d^2\vec{P}_{k\perp}}
=& \sum_i \int\!\! dx\, dy_j dy_k\,
   H_i\Bigl(x,\frac{Q}{\mu},\mu\Bigr)  
  D_{i\to h_j h_k} \Bigl(y_j,y_k,\vec{P}_{j\perp},\vec{P}_{k\perp},\mu\Bigr)
\delta(\xi_j\!-\!x y_j)\delta(\xi_k\!-\!x y_k)\theta(1\!-\!y_j\!-\!y_k)
  \nonumber\\
=& \sum_i \int_{\xi_j+\xi_k}^1 \frac{dx}{x^2}\:
  H_i\Bigl(x,\frac{Q}{\mu},\mu\Bigr) 
  \: D_{i\to h_j h_k} \Bigl(\frac{\xi_j}{x},\frac{\xi_k}{x},
   \vec{P}_{j\perp},\vec{P}_{k\perp},\mu\Bigr)
  \times \Big[ 1 + {\cal O}(\vec{P}_{a\perp}^{\,2}/Q^2)\Big]\,,
\end{align}
where $\sigma_0$ is the Born-level total cross section. Here $H_i$ describe the production of parton $i$ with momentum fraction $x = k^-/q^-$, and are now known to N$^3$LO~\cite{He:2025hin}. The observed momentum fraction of the hadrons are given by $\xi_a = p_{h_a}^-/q^-$. Further below we will exploit the fact that \eq{dihadprod} remains valid when $\Lambda_{\rm QCD}^2\ll \vec{P}_{a\perp}^2$. 
\end{widetext}

Eq.~(\ref{eq:dihadprod}) provides the input needed in \eq{EECdef} when $h_i\ne h_j$. The EEC also has a contribution with $h_i = h_j$, with the two detectors on the same hadron. This contribution comes from the single-hadron production given by
\begin{align}
\label{eq:singlehadprod}
\frac{1}{\sigma_0}\frac{d\sigma^{h X}}{ d\xi}
 =& \sum_i \int dx dy\, H_i\left(x,\frac{Q}{\mu},\mu\right)  D_{i\to h}\left(y,\mu\right)
  \\
&\times\delta(\xi-x y)\theta(1-y)
 \nonumber\\
 =& \sum_i \int_{\xi}^1 \frac{dx}{x}\, H_i\left(x,\frac{Q}{\mu},\mu\right) D_{i\to h}\left(\frac{\xi}{x},\mu\right)\,,
 \nn
\end{align}
with single-hadron FFs
\begin{align}
\label{eq:singlehadQk}
D_{q\to h}(y)=\: &\frac{1}{y}\frac{1}{2N_c}\int \frac{d r^{+}}{2(2 \pi)} e^{i k^- r^+}\\
&\hspace{-1.5cm}\times\sum_X \hspace{-0.5cm} \int\,  \operatorname{Tr}\left[\gamma^- \langle 0|  \psi_q(r^+,0,0_\perp)\left|h; X\right\rangle\left\langle h; X\right| \bar{\psi}_q(0)|0\rangle\right]\,,\nonumber
\end{align}
for quark, and similarly for gluon.
By RG consistency, the evolution equations of the single and dihadron FFs are described by the timelike DGLAP kernel $P^T$, known up to three-loop accuracy~\cite{Mitov:2006ic,Mitov:2006wy,Moch:2007tx,Chen:2020uvt}, as
\begin{align}
\label{eq:RGsingle}
& \frac{dD_{i\to h}\left(\xi, \mu\right)}{d\ln \mu^2} =\sum_{j} \int_{\xi}^1 \frac{dy}{y} D_{j\to h}\left(\frac{\xi}{y}, \mu\right) P^T_{ji}(y)\,,\\
& \frac{dD_{i\to h_k h_l}\left(\xi_k, \xi_l, \vec{P}_{k\perp},\vec{P}_{l\perp} , \mu\right)}{d\ln \mu^2} 
\label{eq:RGdi}
\\
& \quad=\sum_{j} \int_{\xi_k+\xi_l}^1 \frac{dy}{y^2} D_{j\to h_k h_l}\left(\frac{\xi_k}{y},\frac{\xi_l}{y},  \vec{P}_{k\perp},\vec{P}_{l\perp} , \mu\right) P^T_{ji}(y)\,.\nonumber
\end{align}

\emph{\textbf{EEC in the collinear region.}} 
We now connect the results in~\eqs{dihadprod}{singlehadprod} with the $z$ distribution of the EEC in~\eq{EECdef}. The transverse momentum difference between the two hadrons can be expressed in terms of $z$ as
\begin{widetext}
\begin{align}
z =&\frac{1-\cos\theta_{jk}}{2}
= \frac{(\xi_k\vec{P}_{j_\perp}-\xi_j\vec{P}_{k_\perp})^2}{(\xi_j\xi_k)^2Q^2}
   \left(1+ \mathcal{O}\biggl(\frac{\vec{P}^2_{a\perp},M_a^2}{Q^2}\biggr)
  \right)\,.
\end{align}
This allows one to write
\begin{align}
\label{eq:EECfact}
\hspace{-0.25cm}\frac{d\sigma_{\rm EEC}}{dz} \bigg|_{z\ll 1}&= \frac{1}{2^2}\frac{d}{dz}\bigg[\sum_{h_j\neq h_k}\int d\xi_j d^2\vec{P}_{j\perp}d\xi_k d^2\vec{P}_{k\perp}\xi_j \xi_k\,\Theta\left(z-\frac{(\xi_k\vec{P}_{j_\perp}-\xi_j\vec{P}_{k_\perp})^2}{(\xi_j\xi_k)^2Q^2}
\right)\frac{d\sigma^{h_j h_kX}}{ d\xi_j d\xi_k d^2\vec{P}_{j\perp}d^2\vec{P}_{k\perp}}\nonumber\\
& \hspace{2cm}+ \sum_h \int d\xi\, \xi^2\, \frac{d\sigma^{hX}}{ d\xi}\Theta(z)\bigg] \nonumber\\
&= \sigma_0\frac{d}{dz}\left[\sum_i\int dx\, x^2\,H_i\left(x,\frac{Q}{\mu},\mu\right) J_{i}(x^2Q^2 z,\mu)\right]\,,
\end{align}
where the EEC jet function is expressed in terms of fragmentation functions
\begin{align}
\label{eq:JetFcnDihad}
J_i (x^2 Q^2z,\mu) =& \frac{1}{2^2}\sum_{h_j\neq h_k}\int \frac{d^2\vec{P}_{j\perp}}{(2\pi)^3}\frac{dy_j }{2y_j} \int \frac{d^2\vec{P}_{k\perp}}{(2\pi)^3}\frac{dy_k }{2y_k}\,  \,\Theta\left(x^2Q^2 z-
\frac{(y_k\vec{P}_{j_\perp}-y_j\vec{P}_{k_\perp})^2}{(y_jy_k)^2}
\right)\nonumber\\
&\times y_j y_k (2\pi)^6 (2y_j 2y_k) \,D_{i\to h_j h_k}\left(y_j,y_k,\vec{P}_{j\perp},\vec{P}_{k\perp},\mu\right)+ \frac{1}{2^2}\sum_h \int dy\,y^2  D_{i\to h}(y,\mu)\,\Theta(x^2Q^2 z)\,.
\end{align}
The overall $\frac{1}{2^2}$ comes from normalizing by $Q^2$ in the denominator of~\eq{EECdef}.
This formula for the EEC jet function applies even when $x^2 Q^2 z \sim \Lambda_{\rm QCD}^2$ and makes explicit that the jet function only depends on the $x^2 Q^2 z$ combination.  

Using the bare operator definitions of the single- and di-hadron FFs  in~\eqs{singlehadQk}{dihadQk},
we can obtain a more familiar expression for $J$. 
First, we shift the coordinate dependence of the quark field $\psi_q$ and carry out the integration to produce the momentum-conserving delta function. We then redefine $X$ to include the explicit hadronic states in the FFs, identifying $h_j h_k X \to X$ in the di-hadron case and and $h X \to X$ in the single-hadron case. Finally, we note that the single-hadron sum $\sum_h$ coincides with the $h_j = h_k$ sector of the double sum $\sum_{h_j,h_k}$. Altogether we then obtain after renormalization indicated by the subscript $\mu$
\begin{align}
\label{eq:jetfcn}
J_q(x^2 Q^2z,\mu) =& \frac{1}{2N_c} \sum_X\hspace{-0.5cm}\int \: \sum_{h_j\, h_k\in X}\frac{E_j E_k}{x^2Q^2} \,\Theta\left( x^2Q^2 z - 
\frac{(y_k\vec{P}_{j_\perp}-y_j\vec{P}_{k_\perp})^2}{(y_jy_k)^2}
\right) \nonumber\\
&\times(2\pi)^3\delta(x Q/\sqrt{2}-p_X^-)\delta^2(\vec{p}_{X,\perp})\operatorname{Tr}\left[\gamma^- \langle 0|  \psi_q(0)\left|X\right\rangle\left\langle X\right| \bar{\psi}_q(0)|0\rangle\right]_\mu\,,
\end{align}
\end{widetext}
and similarly for the gluon jet.  
What we have achieved is an extension of the factorization to the full collinear limit, including the region $Q^2z \sim \Lambda_{\rm QCD}^2$, where the jet function can be written as transverse-momentum-sensitive dihadron FFs as~\eq{JetFcnDihad}. Thus, we have identified the universal nonperturbative functions that are sensitive to the EEC's confinement transition!

\emph{\textbf{Factorization in OPE region}}
The result in \eq{jetfcn} reproduces the collinear jet function in the OPE region $\Lambda_{\rm QCD}^2/Q^2\ll z\ll 1$ presented in~\cite{Dixon:2019uzg}, where the jet function can be perturbatively computed by taking $h_j h_k$ to be partons. 
It is interesting to see if this parton level result can be derived from the $\Lambda_{\rm QCD}^2\ll Q^2 z$ limit of \eq{JetFcnDihad}. 
Based on standard collinear factorization
the transverse-momentum-sensitive dihadron FFs in the limit $\Lambda_{\rm QCD}^2\ll \vec{P}_{a\perp}^2$ will refactorize as 
\begin{widetext}
\begin{align}
\label{eq:dihadrefact}
\hspace{-0.6cm}D_{i\to h_j h_k} (y_j,y_k, \vec{P}_{h_j\perp},\vec{P}_{h_k\perp},\mu)
 =& \: \sum_{j,k} \int_{y_j}^1\! \frac{dw_j}{w_j}\! \int_{y_k}^1\! \frac{dw_k}{w_k}
  \, C_{i\to jk}\left(w_j,w_k,\vec{P}_{j\perp},\vec{P}_{k\perp},\mu,\mu'\right)
  \, D_{j\to h_j}\left(\frac{y_j}{w_j},\mu'\right)
   D_{k\to h_k}\left(\frac{y_k}{w_k},\mu'\right)
   \nonumber\\
&+\sum_j \int_{y_j+y_k}^1 \frac{dw}{w^2} C_{i\to j}(w,\vec{P}_{j\perp},\mu,\mu') D_{j\to h_j h_k}\left(\frac{y_j}{w},\frac{y_k}{w},\mu'\right)\delta^2\left(\vec{P}_{h_j\perp}-\frac{y_j}{y_k}\vec{P}_{h_k\perp}\right)\,.
\end{align}    
\end{widetext}
In the coefficient $C_{i\to jk}$, the transverse‑momentum argument, $\vec P_{j\perp}=\frac{w_j}{y_j}\,\vec P_{h_j\perp}$ denotes the transverse momentum of parton $j$ in terms of the transverse momentum $\vec P_{h_j\perp}$ of the hadron into which it fragments.
Here $\mu'$ is the refactorization scale between the partonic coefficients $C$ and collinear FFs. Note that for $Q^2 z \gg \Lambda_{\rm QCD}^2$ the contact terms in the last line of \eq{dihadrefact} can still be present. Since angular scales of size $Q^2 z \sim \Lambda_{\rm QCD}^2$ are no longer tracked, the contact term is given by purely collinear dihadron FFs without explicit transverse momentum dependence. We believe this is the first presentation of the refactorization  in~\eq{dihadrefact}, including its contact terms, although a similar expression for the first term can be found~\cite{Ju:2025jxi,Ceccopieri:2007ip,Zhou:2011ba}.  

The contact term is important for IRC safety of the cumulative energy correlators in the OPE region.
After taking moments as in~\eq{jetfcn}, by the momentum sum rule of the single-hadron FFs the first term in \eq{dihadrefact} becomes $\sum_{jk} \int dw_j dw_k w_j w_k\, C_{i\to jk}(w_j,w_k,\vec{P}_{j\perp},\vec{P}_{k\perp},\mu)$ (up to transverse momentum integrals we suppress for convenience), where $C_{i\to jk}$ is given by the same matrix element as~\eq{dihadQk} with hadronic states replaced by partonic states.  
Since the sum rule is independent of $\mu'$, the $\mu'$ in $C$ disappears in this moment. 
Taking the same moment of the last term in~\eq{dihadrefact} and combining with the single-inclusive hadron contribution in~\eq{JetFcnDihad}, we find that the full contact term is
\begin{align}
&\sum_j \int_0^1 dw\,w^2\, C_{i\to j}(w,\mu,\mu')\bigg[\sum_h\int dy\, y^2 D_{j\to h}(y,\mu')\nonumber\\
&+ \sum_{h_j\neq h_k} \int dy_j\,dy_k \,y_j\,y_k\,D_{j\to h_j h_k}(y_j,y_k,\mu')\bigg]\theta(x^2Q^2z) \nonumber\\
& \hspace{1cm}=\sum_j \int_0^1 dw\,w^2\, C_{i\to j}(w,\mu)\,\theta(x^2Q^2z)
  \,,
\end{align}
where we used the momentum sum rule of the collinear single- and di-hadron FFs. The coefficients $C_{i\to j}$ is given by the same matrix element as~\eq{singlehadQk} with hadronic states replaced by partonic states. We thus obtain the jet function in~\eq{jetfcn} with $h_j,h_k$ replaced with partonic states, as expected in the OPE region. This is consistent with the expectation that EEC is an IRC safe observable and that IR singularities cancel between the analogous two terms in partonic calculations. Exclusive measurements on hadronic states can introduce sensitivity to various collinear FFs organized in terms of moments of track functions as noted in~\cite{Jaarsma:2023ell,Lee:2023npz,Lee:2023tkr,Chen:2020vvp} and this is apparent from our derivation here.

\emph{\textbf{Predicting the $Q$ dependence of $d\sigma_{\rm EEC}/dz$ for the Confinement Transition}}  
Although $J$ is nonperturbative in the $Q^2z \sim \Lambda_{\rm QCD}^2$ region, the jet function given in terms of dihadron FFs has a perturbative RG evolution structure
\begin{align}
\label{eq:jetRG}
& \frac{dJ_{i}(Q^2 z,\mu)}{d\ln \mu^2} =\sum_{j} \int_0^1 dy\, y^2 J_{j}(y^2Q^2 z,\mu) P^T_{ji}(y)\,,
\end{align}
which follows from the result in~\eq{JetFcnDihad} and the RG evolution of the FFs in~\eq{RGsingle} and~\eq{RGdi}.
Eq.~(\ref{eq:jetRG}) is also identical to the RG evolution of the perturbative collinear jet function in the OPE region~\cite{Dixon:2019uzg}. 

\begin{figure}[!t]
\begin{center}
\includegraphics[width=0.48\textwidth]{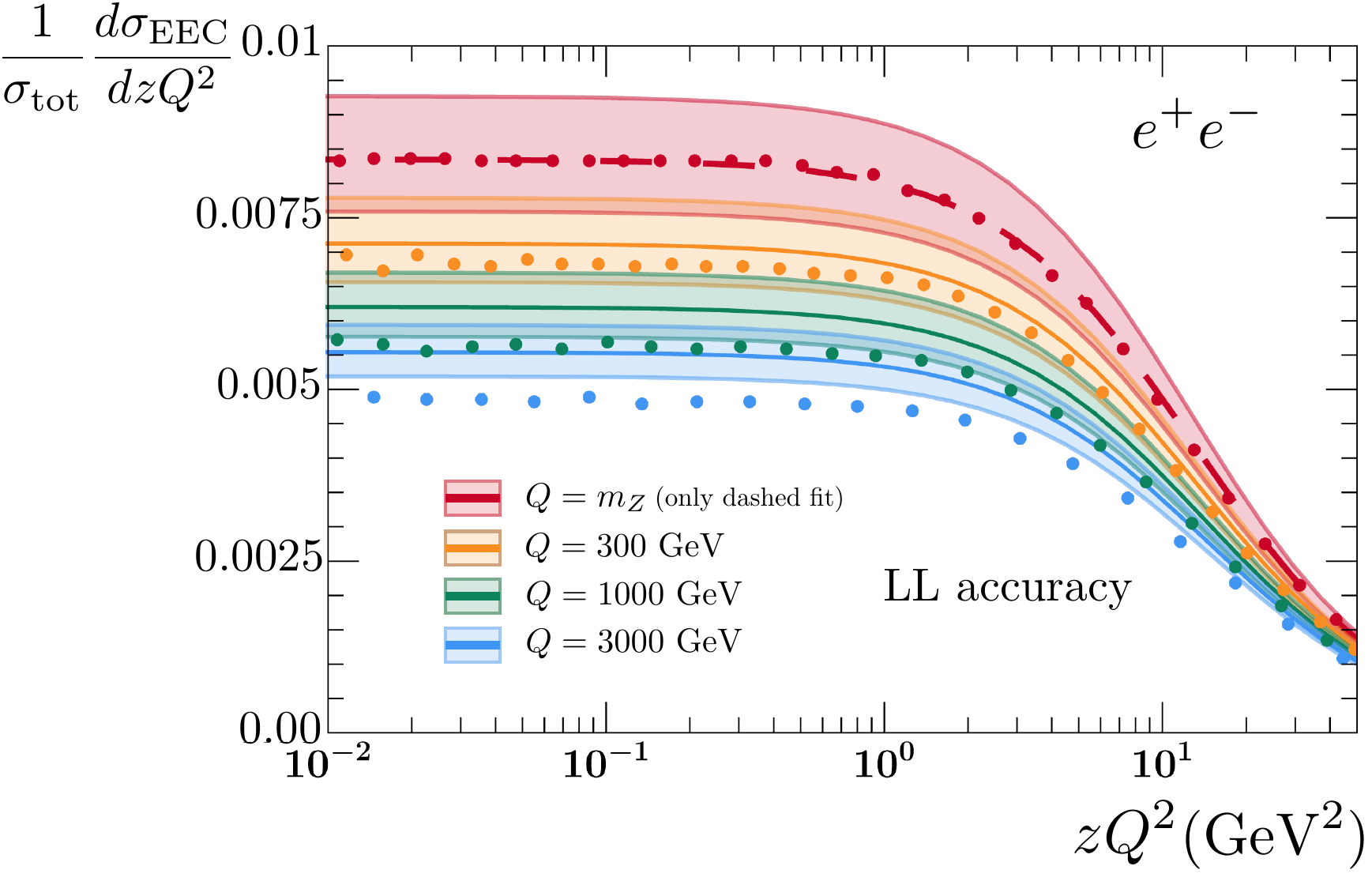}
\includegraphics[width=0.48\textwidth]{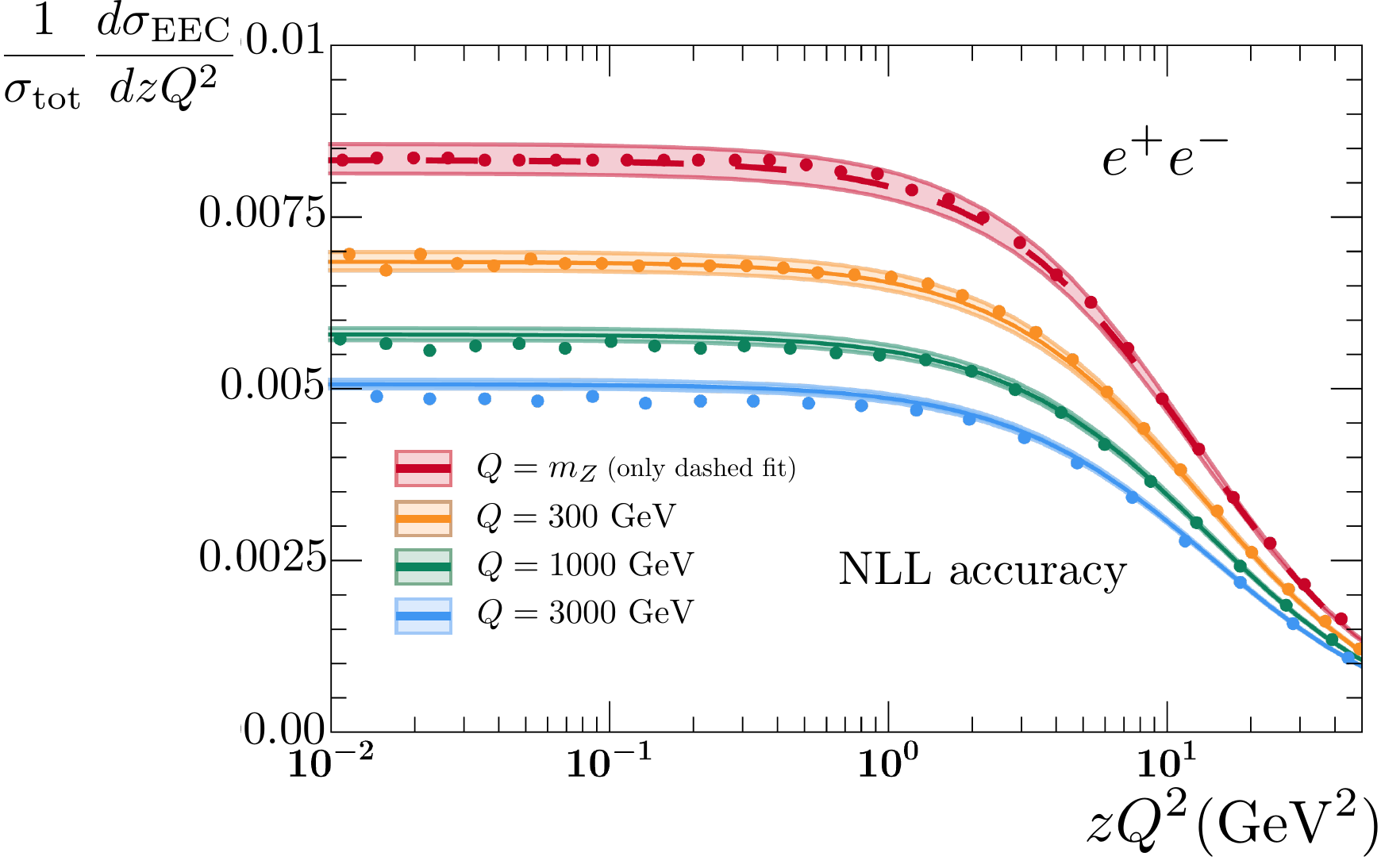}
\end{center}
\caption{Leading logarithmic (top) and next-to-leading logarithmic (bottom) predictions of EEC as a function of $z Q^2$ in the $z Q^2\sim \Lambda_{\rm QCD}^2$ region. Only the central curve (dashed) at $Q=m_Z$ is fit, while rest of the curves are derived according to the perturbative evolution from~\eq{jetRG}.}
\label{fig:Qdep}
\end{figure} 
We now follow the standard program of extracting the nonperturbative $J_i$ at one scale and evolving them to others to test factorization. At some $\Lambda_{\rm trans}^2 \gtrsim \Lambda_{\rm QCD}^2$ we have the striking confinement transition. When $zQ^2 \gtrsim \Lambda_{\rm trans}^2$, $d\sigma_{\rm EEC}/dz\sim z^{-1-\delta}$ with anomalous dimensions $\delta\leq0$ and power corrections~\cite{Korchemsky:1999kt,Schindler:2023cww,Lee:2024esz,Chen:2024nyc} $\delta>0$ both contributing to the scaling. On the other hand, when $zQ^2 \lesssim \Lambda_{\rm trans}^2$, $d\sigma_{\rm EEC}/dz\sim z^0$, ie.~flat due to confinement. This motivates using the following model to fit the jet function,\footnote{One may wonder whether it is necessary to fit in the region where $Q\sqrt{z} > \Lambda_{\rm trans}$, since $J_i$ in the OPE region can be computed. However, this description becomes worse as we approach $\Lambda_{\rm trans}$. Since the OPE result follows the same RGE, there is no harm from extending our fit into the OPE region, as long as its consistent with perturbative calculations. One can of course match more systematically between the two regions.}
\begin{align}
\label{eq:inputJet}
J_{i}(Q^2 z,\mu_0)  =\int^{Q^2z}_0 d\nu\,\frac{N_i}{1+(\frac{\nu}{\Lambda_i})^{1+\delta_i}} \,,
\end{align}
where $N_i$ is a normalization parameter, $\Lambda_i$ denotes a transition scale, and $-1-\delta_i$ is the scaling exponent when $Q^2z > \Lambda_i$. 
Note that we did not exploit the (di)hadron‐FF formula in~\eq{JetFcnDihad} to motivate \eq{inputJet}. Instead, we apply~\eq{JetFcnDihad} in reverse to provide deeper physical insight into the dihadron FFs through the confinement transition exhibited by the EEC. 
In the flat confining region this predicts that a weighted integral (``sum-rule") over the (di)hadron fragmentation function predicts a remarkable free streaming behavior from the cloud of hadrons produced via the confining fragmentation process. 
The analog in the OPE scaling region predicts that this weighted integral over the fragmentation functions has a perturbatively calculable dependence on $z Q^2$, which  eventually transitions into the free streaming hadrons as it becomes confined.

Using \eq{inputJet} we extract the quark and gluon jet functions at $\mu_0 = 10\,\text{GeV}$ using Pythia simulations of $e^+e^-\to \gamma^*/Z\to q\bar{q}$ and $H\to gg$ with $Q=m_H=91.2~\mathrm{GeV}=m_Z$, by evolving $J_i$ to $\mu=m_Z$ and combining it with the LO coefficients $H_i$. 
We chose this $\mu_0$ since the transition begins at $\sqrt{z} Q \sim 10\,$GeV. 
In \fig{Qdep} (upper panel) we show the results from LL RG evolution
(using Runge-Kutta), 
where the dashed $Q=m_Z$ curve is the starting input. We use $J_i$ at $\mu = Q, Q/2, 2Q$ to obtain an estimate for the perturbative uncertainties, and observe that the evolution agrees with the general trend seen also in MC, though the uncertainties are fairly large.
In~\fig{Qdep} (lower panel), we repeat the exercise with NLO $H_i$, and carrying out NLL RG evolution.  We note a dramatic reduction in the perturbative uncertainty, and better agreement with Pythia at all $Q$ values. This gives a useful cross check on the validity of our results. 

Recent work in Refs.~\cite{Liu:2024lxy,Barata:2024wsu} have also explored the $Q$ dependence of $d\sigma_{\rm EEC}/dz$ in the $z\ll 1$ confinement transition region,
but with TMD-factorization–inspired models rather than first-principles QCD.
In~\cite{Liu:2024lxy} $d\sigma_{\rm EEC}/dz$ was modeled by the TMD distribution structure with a nonperturbative Sudakov exponent, and the $Q$ 
dependence was predicted by Collins-Soper evolution within the Sudakov.
In~\cite{Barata:2024wsu} they exploited collinear factorization, and then TMD modeling for $J_i$ with a (different) nonperturbative Sudakov factor in the confinement transition region. They predicted the $Q$ dependence by evolving their jet function at LL accuracy according to~\eq{jetRG}, and observed the same trend seen in our \fig{Qdep} (upper panel).
Our work puts such tests of the $Q$ dependence of the confinement transition on a firm theoretical footing.

\emph{\textbf{Conclusions and Outlook.}} 
In this letter, we proved the collinear factorization of EEC in the hadron transition region $Q^2z \sim \Lambda_{\rm QCD}^2$ and determined its universal nonperturbative jet functions 
in terms of transverse-momentum-sensitive dihadron fragmentation functions.
This enables predictions of the $Q$ dependence through RG evolution, and a QCD based interpretation of the confinement transition. We also provide 
a factorization of dihadron FFs for $\Lambda_{\rm QCD}^2\ll \vec{P}_{a\perp}^2\ll Q^2$ which is consistent with the collinear factorization of the EEC in the OPE region~\cite{Dixon:2019uzg}.

Our work provides exciting connections between the fields of energy correlators 
and multi-hadron production,
providing a powerful tool for phenomenological applications. It is straightforward to generalize our results to higher-point energy correlators~\cite{Chicherin:2024ifn,Chen:2019bpb,Chen:2020vvp,Yang:2022tgm,Chen:2023zlx}, and when $\vec{P}_{a\perp}^2\sim \Lambda_{\rm QCD}^2$ we expect the nonperturbative dynamics to be described by higher multiplicity transverse-momentum-sensitive multi-hadron fragmentation functions. These can also be measured with different quantum numbers~\cite{Lee:2023npz,Lee:2023tkr,Jaarsma:2023ell,Chen:2024nfl,vonKuk:2025kbv,Mantysaari:2025mht}, such as spin, charge, and heavy-flavor number,
and it would be interesting to track how this affects the nonperturbative confinement transition. It would also be interesting to calculate the EEC jet function with chiral perturbation theory~\cite{Copeland:2024cgq,Copeland:2024wwm,Chen:2001nb}. 
Finally, applying this formalism to probe jet~\cite{Generet:2025vth,Craft:2022kdo,Lee:2022uwt,Komiske:2022enw,Andres:2022ovj,Barata:2023bhh,Holguin:2022epo,Devereaux:2023vjz} and nuclear~\cite{Liu:2022wop,Liu:2023aqb,Cao:2023qat} substructures in the confinement transition region will be very interesting.

{\it Acknowledgements.}---We thank Marston Copeland, Zhongbo Kang, Johannes Michel, Ian Moult, Duff Neill, and Fanyi Zhao for helpful discussions. This work was supported by the U.S. Department of Energy, Office of Science, Office of Nuclear Physics from DE-SC0011090. I.S. was also supported in part by the Simons Foundation through the Investigator grant 327942.


\bibliography{EEC_ref.bib}{}
\bibliographystyle{apsrev4-1}
\newpage
\onecolumngrid
\newpage

\end{document}